\documentclass[aps,prl,twocolumn]{revtex4-2}
\usepackage[english]{babel}
\usepackage{graphicx}
\usepackage{latexsym}
\usepackage{float}

\begin{document}
\title{Single file motion of robot swarms}

\author{Laciel Alonso-Llanes}

\author{Angel Garcimart\'in}
\email{angel@unav.es}

\author{Iker Zuriguel}
\affiliation{Departamento de F\'{i}sica, Facultad de Ciencias, Universidad de Navarra, E-31080 Pamplona, Spain}

\begin{abstract}
We present experimental results on the single file motion of a group of robots interacting with each other through position sensors. We successfully replicate the fundamental diagram typical of these systems, with a transition from free flow to congested traffic as the density of the system increases. In the latter scenario we also observe the characteristic stop-and-go waves. The unique advantages of this novel system, such as experimental stability and repeatability, allow for extended experimental runs, facilitating a comprehensive statistical analysis of the global dynamics. Above a certain density, we observe a divergence of the average jam duration and the average number of robots involved in it. This discovery enables us to precisely identify another transition: from congested intermittent flow (for intermediate densities) to a totally congested scenario for high densities. Beyond this finding, the present work demonstrates the suitability of robot swarms to model complex behaviors in many particle systems.
\end{abstract}
\maketitle

In recent years, swarm robotics has emerged as a fascinating field where the goal is the design and control of robot ensembles operating as a coordinated system \cite{swarmrobotics1,swarmrobotics2,swarmrobotics3,Hamann}. These robots are typically simple and small, and they interact with each other and their environment to achieve a common goal. Despite their limited individual capabilities, they can be programmed to collaboratively perform complex tasks such as searching for objects, exploring unknown environments, or transporting objects. Robot swarms also provide a unique platform for studying collective behavior. For instance, \emph{kilobots} \cite{kilobots,kilobots2,kilobots3}, characterized by their compact size, straightforward design, and controllability through an open-source platform, stand out as excellent agents for conducting large-scale studies involving hundreds of units. Importantly, two distinct approaches can be adopted in the investigations of collective motion with robot swarms: i) a pragmatic approach primarily focused on the task to be performed, where the robot rules are fine-tuned to optimize the collective performance; and ii) a bottom-top approach in which the motion and interaction rules among the agents are established and the emerging macroscopic dynamics are analysed. We took the latter one, which resembles the traditional approach implemented in the science of complex systems such as active matter, pedestrian crowd dynamics and collective animal behavior.

\begin{figure}
\includegraphics[width=0.95\columnwidth]{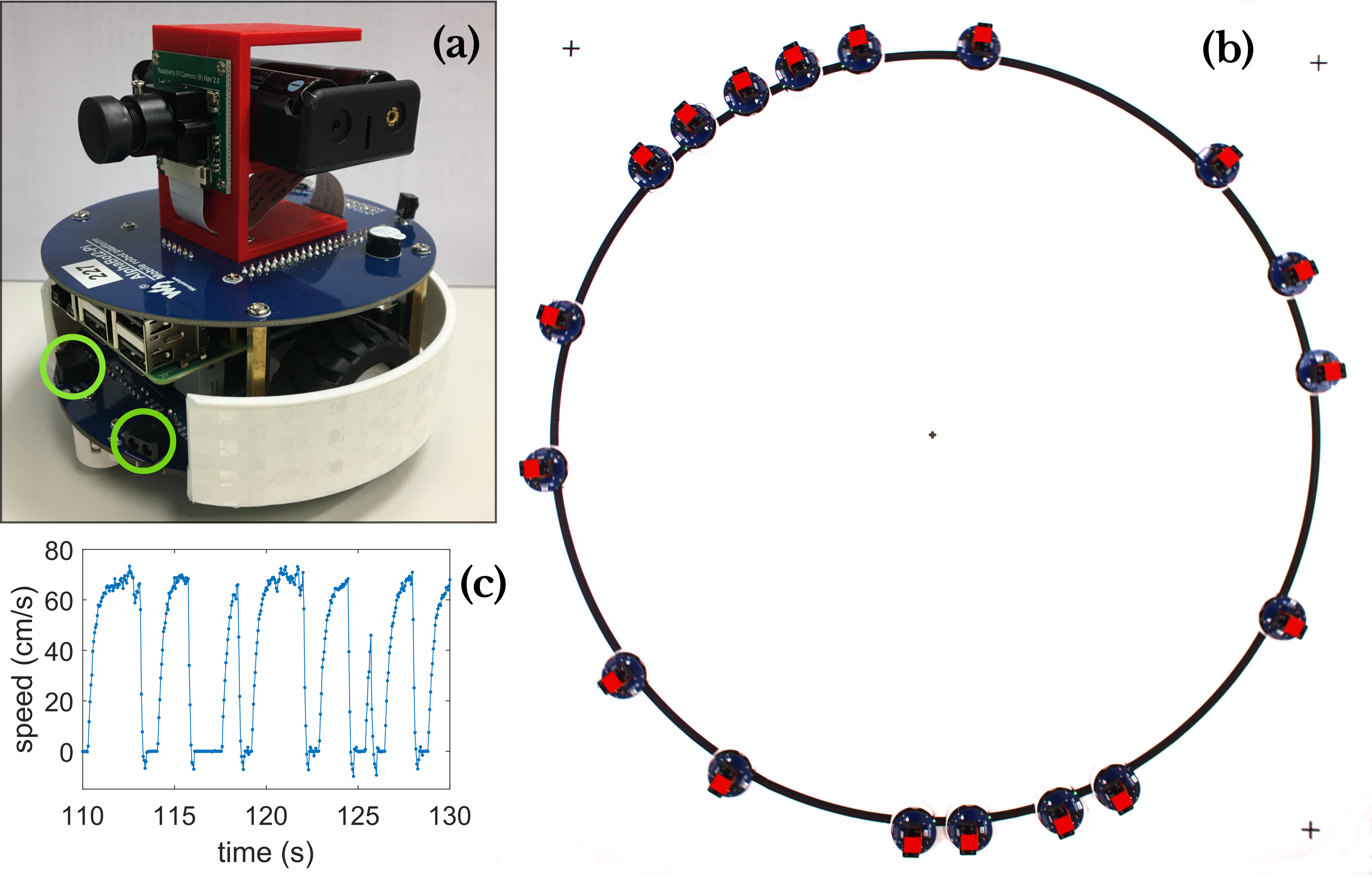}
\caption{\label{fig:exp} (a) Photograph of a robot. Green circles mark the infrared proximity sensors. (b) Snapshot of 18 robots on the circular lane; a jam of five robots can be seen at approximately 11 o’clock. (c) Single-robot speed over a 20-second interval in an experiment with $N=20$ robots and $V_{max}=69$ cm/s.}
\end{figure}

We have used an assembly of palm-sized robots that incorporate an on-board nanocomputer, which allows a high degree of control and flexibility on the programmed interactions --significantly greater than the simpler kilobots, for instance. The trade-off is a reduction in the number of agents that make up the swarm, with a hundred being a reasonable upper limit. Weighing the advantages against the disadvantages, we propose this robot swarm as a suitable framework to investigate the emergence of collective behaviors in active matter such as self-organization in confined geometries \cite{confinement}, clogging in bottleneck flows \cite{cloggingbugs}, segregation driven by counterflow \cite{Chowdhury}, and any other scenario where the role of inertia or physical contacts between agents is significant.

As a starting point, we have chosen investigating the motion in a single file experiment. The main reason for this choice lies in the simplicity of the system, but also the extensive number of experiments developed on this topic \cite{Taloni}, especially in fields such as vehicular traffic \cite{Sugiyama,Zhang} and pedestrian dynamics \cite{Seyfried,Cao,Lu}. Additionally, there exists a theoretical framework \cite{NagelSchreckenberg,Krauss,Schadschneiderrevisited,Nagatani} against which we can contrast our findings. Essentially, the main features observed in single file motion include: i) a transition from free flow to a congested flow as density (number of agents per unit length) increases; and ii) the spontaneous emergence of stop-and-go waves travelling backwards. Also, the single file geometry has been largely used to investigate the dependence of the flow rate on the density in pedestrian dynamics, a representation known as the fundamental diagram \cite{Vanumu,ChowdhuryVEHICULAR,Parisi}. Perhaps the main drawback of the experiments conducted thus far is the limitation in the duration of the experimental runs, along with a lack of control and stability over the experimental conditions. Remark that pedestrian dynamics and vehicular traffic are affected by human behavior changing over time, due to factors such as tiredness or boredom. In other active systems, maintaining experimental conditions unaltered can also be a challenging task. It is precisely on this point that experiments with robots can be extremely useful, since their behavior remains constant over time with the only limitation being the duration of the batteries (more than one hour in our case). Therefore, this system enables the implementation of long and steady experimental runs, a critical factor in revealing certain distinctive statistical features of the collective dynamics.

Our robots are 11 cm in diameter (Fig.~\ref{fig:exp}\textbf{a}) and are constructed using the Alphabot2 platform, a kit manufactured by Waveshare, which incorporates a Raspberry Pi nanocomputer. They are driven by two wheels propelled by DC motors and are equipped with various elements enabling control over interactions with other agents and the environment. Infrared (IR) sensors at the base can detect a dark line and follow it by regulating the power delivered to each wheel; a PID loop is used to this end. Two infrared sensors at the mid-front detect obstacles or other robots within a tunable distance; for this work, we have set the distance to 15 cm. The robot speed can be adjusted by changing the motor duty cycle; here we tested speeds ranging from about 20 to 80 cm/s. Remarkably, the speed exhibits slight variations from one robot to another, with an interquartilic range of about 10\% for a given duty cycle. On top of the standard equipment, our robots have been upgraded by including a red support for the camera (also used for tracking robots), batteries with greater capacity, and a white, reflective shell. The on-board computers are connected to an external computer via Wi-Fi, allowing remote access and operation. For detailed information about the robot features, see the Supplemental Material (SM) \cite{SM}.

\begin{figure}
\includegraphics[width=\columnwidth]{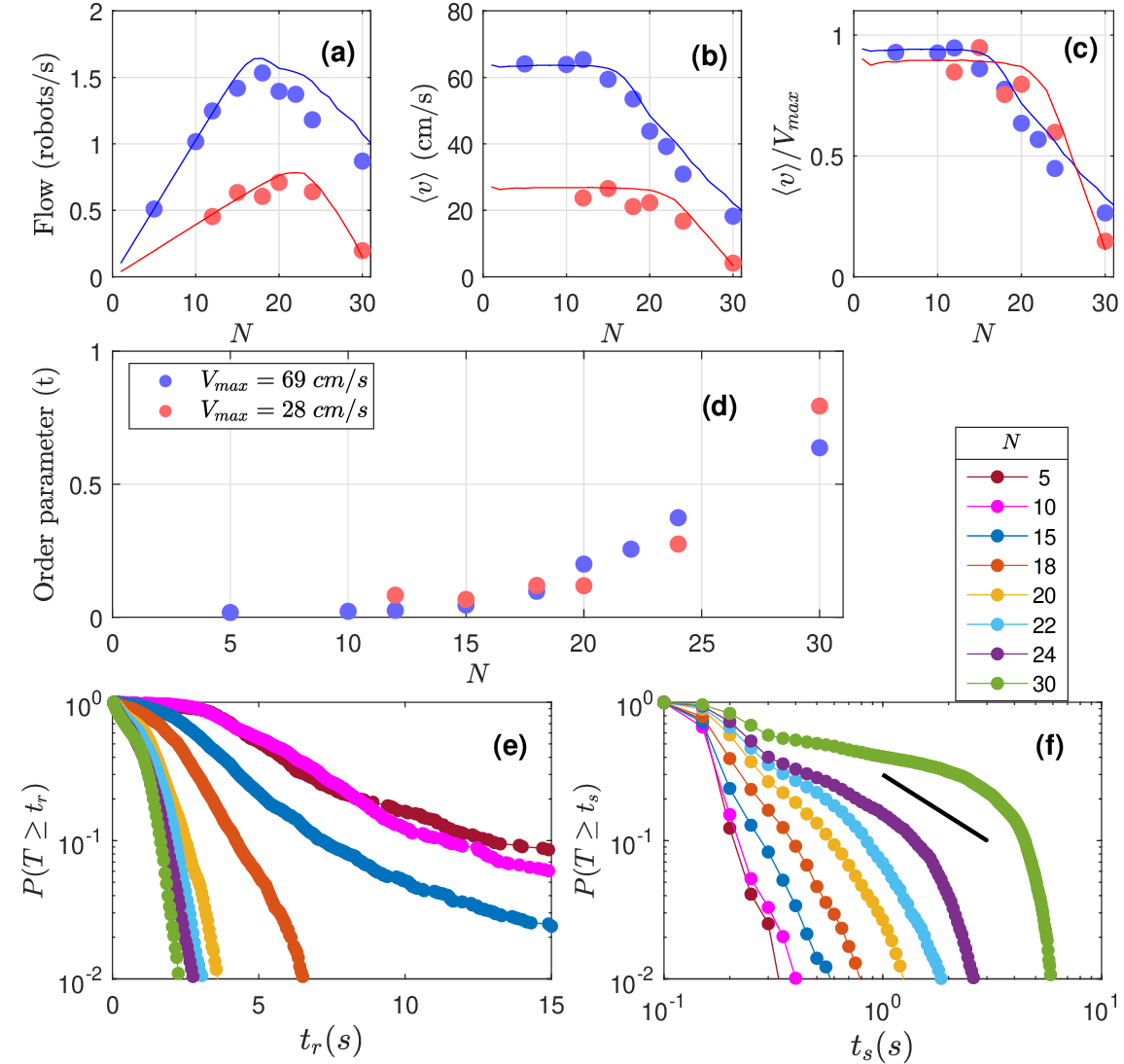} \par
\caption{\label{fig:dia} (a) Average flow rate versus the number of robots in the system (N) for two different robot speeds ($V_{max}=28$ cm/s and $V_{max}=69$ cm/s, as indicated in the legend of d). Solid lines are results of numerical simulations. (b) Average speed $\langle v \rangle$ as a function of the number of robots $N$. (c) Rescaled speed (\emph{i.e.}, the average speed divided by robot speed $V_{max}$), as a function of $N$. (d) Order parameter $t$ as a function of N. (e) Log-lin survival function of the time lapse that an individual robot is running $t_r$, for scenarios with $V_{max}=69$ cm/s and different $N$ (as indicated in the legend above panel (f)). (f) Survival function of the time lapse that an individual robot is stopped $t_s$ in log-log scale. The solid line has a slope of $-1$, hence corresponding to a power-law exponent of $-2$. }
\end{figure}

At the beginning of an experimental run, a set of $N$ robots is arranged at equal intervals atop a 2 m diameter circular black line (Fig.~\ref{fig:exp}\textbf{b}). All robots receive the same configuration defining operational parameters, and they are set into motion simultaneously. Subsequently, robots follow the line at the prescribed speed and stop whenever they detect another robot in front of them. Then, as soon as the sensors do not detect anything in front, the robot resumes motion. Therefore, the nominal speed can be 0 (robot stopped) or $V_{max}$ (the `free speed'). Crucially \cite{SlowToStart1,SlowToStart2}, the braking is more abrupt than the acceleration, as evidenced in the speed vs. time graph of a single robot represented in Fig.~\ref{fig:exp}\textbf{c}. It takes approximately 0.2 s to reduce $V_{max}$ by one half, while about 0.5 s are needed to accelerate from 0 to $V_{max}/2$. The motion of the robots is recorded with a video camera from above with a resolution of $2600 \times 2600$ pixels at 20 frames per second (see video in \cite{SM}). We register several runs of 3 minutes for each set of conditions. From these, we obtain the robot position by analysing each frame and detecting the center of red squares, hence obtaining a file with the robot number, time, and position, with an accuracy of $\pm 0.05$ s in time and $\pm 2 $ mm in position.

We first show the dependence of macroscopic quantities, such as flow and velocity, on the number of robots $N$ in the system. In Fig.~\ref{fig:dia}\textbf{a} we observe the existence of a region of free flow where the flow grows linearly with $N$, and then a congested region where the flow decreases with $N$. This observation holds for all nominal free velocities, $V_{max}$, although only two (28 cm/s and 69 cm/s) are shown in the plot. The existence of these two regimes is also apparent in the representation of the average robot velocity ($\langle v \rangle$) versus $N$. In the free flow region, $\langle v \rangle$ remains constant and does not depend on $N$, whereas in the congested region it noticeably decreases with $N$. The transition seems to occur in the range $15 < N < 18$ irrespective of $V_{max}$. This characteristic is underscored in Fig.~\ref{fig:dia}\textbf{c} when normalizing the average velocity to $V_{max}$. A standard way to visualize this transition is by computing the fraction of time that the agents are stopped $t= \frac{t_s}{t_s+t_r}$, where $t_s$ and $t_r$ are the times that the robots are stopped and running respectively. Results (Fig.~\ref{fig:dia}\textbf{d}) show that $t$ is close to 0 up to $15<N<18$ and increases beyond this point, confirming the existence of the transition.

Taking a further step, we investigated the features of the collective motion by examining the statistics of running and stop intervals through survival functions (i.e., the probability of finding an interval lasting longer than a given time). To this end, we focused on the case $V_{max}$ = 69 cm/s (similar outcomes are obtained for other velocities). The running intervals (Fig.~\ref{fig:dia}\textbf{e}) exhibit a clear change in trend at $15<N<18$; for larger $N$, the distributions appear to follow an exponential pattern, while the tails broaden for smaller $N$ (free flow region) due to robots experiencing unboundedly long $t_r$. Similarly, the distributions of $t_s$ (Fig.~\ref{fig:dia}\textbf{f}) show an increase in stopped time as $N$ grows. Interestingly, for $N=30$ the distribution seems compatible with a power-law with an exponent smaller than 2 (indicated by the black line as a reference), at least for a small range of $0.3<t_s<3$  seconds. If confirmed, this observation will have significant consequences, as a power-law distribution with an exponent smaller than two implies a lack of convergence of its first moment (i.e., the average) \cite{Clauset}. This suggests the possible existence of another transition (for systems with very high densities) to a region with different behavior, where the time a robot remains stopped diverges. Unfortunately, we cannot verify if this power-law tail extends over a broader range of $t_s$ values, as there is an intrinsic temporal cutoff determined by the system size and number of robots: a robot cannot be stopped longer than the time it takes for all robots in front of it to start moving. In other words, despite the analysis of stopping times hints at the existence of a transition to a fully congested state when $N\approx30$, this parameter does not appear to be the most suitable for drawing a definitive conclusion.

\begin{figure}
\includegraphics[width=\columnwidth]{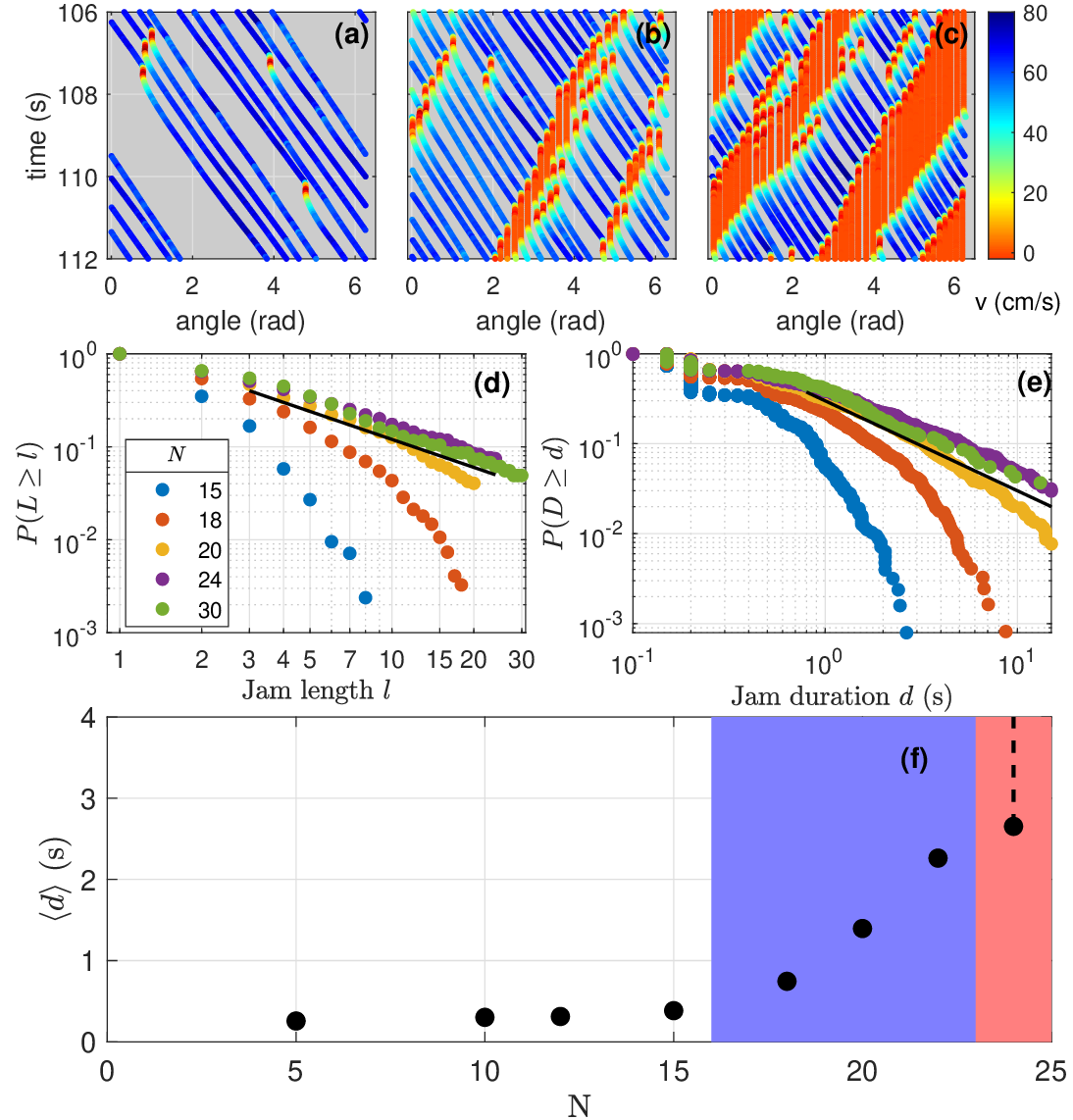}
\caption{\label{fig:survival} (a-c) Spatiotemporal diagrams depicting six seconds of experimental results with $V_{max}=69$ cm/s and $N$=10, 20, and 30 robots, respectively. Speeds are color-coded as shown in the color bar. (d) Survival function of the jam length ($l$) for $V_{max}=69$ cm/s and different numbers of robots ($N$), as indicated in the legend. Note the logarithmic scale. (e) Survival function in logarithmic scale for the jam duration $d$. Solid lines serve as guides to the eye, illustrating an exponent of $-1$ which to power law tails with an exponent of $-2$ in the probability density. (f) Average jam duration $\langle d \rangle$ as a function of $N$. The point corresponding to $N=24$ represents the average of the registered data; remark, however, that this value does not accurately reflect convergence of the first moment, as $\langle d \rangle$ grows unboundedly with the measuring time. This is indicated with the dashed line.}
\end{figure}

Motivated by this discovery, we focused on jam formation and the characteristics of stop-and-go waves, typical of these systems. In Fig.~\ref{fig:survival}\textbf{a-c}, we present a portion of the spatiotemporal diagrams constructed with the angular position of each robot (x-axis) versus time (increasing downwards on the y-axis). When $N=10$ (Fig.~\ref{fig:survival}\textbf{a}), the robots (each represented by a line) move towards the right (increasing values of angle) with short jams appearing only occasionally (red segments). As $N$ increases (Fig.~\ref{fig:survival}\textbf{b}), jams become more frequent, and they clearly propagate backwards (from right to left as time increases). Finally, for $N=30$ (Fig.~\ref{fig:survival}\textbf{c}), jams increase in size (number of robots involved) and live longer. The jammed regions (red segments) seem to percolate under these conditions. From this visual representation, we proceeded to obtain the survival functions of jam lengths ($l$, denoting the number of robots involved in the jam) for different values of $N$ (Fig.~\ref{fig:survival}\textbf{d}). Note that these plots are only meaningful when jams involving more than a few robots are formed (i.e., for $N\geq15$). Interestingly, we observe that the distributions for $N=24$ and $N=30$ are compatible with power law decays with an exponent smaller than 2. As mentioned earlier, this implies that average jam lengths grow unboundedly with the measuring time window, suggesting a state of total congestion. But once again we face the problem of distributions spanning less than one order of magnitude (at most, from $l=3$ to $l=30$). This limitation is, of course, imposed by the number of robots in the ensemble.

\begin{figure}
\includegraphics[width=\columnwidth]{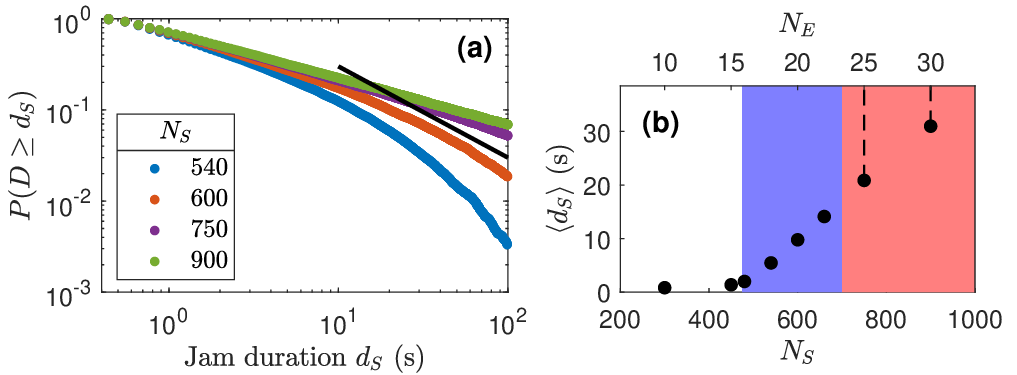}
\caption{\label{fig:simuls} (a) Survival function for jam duration in simulations ($d_S$) for different number of agents ($N_S$). The solid line serves as a guide to the eye, illustrating an exponent of -1 in the survival, corresponding to power laws with an exponent of -2. Note the logarithmic scale. (b) Average jam duration $\langle d_S \rangle$ as a function of $N_S$. The top axis indicates the number of agents ($N_E$) corresponding to systems with equal density and the size of the experimental scenario.}
\end{figure}

A way to overcome this restriction is to examine the statistics of jam duration, a variable constrained only by the length of the experimental runs. Once again, the survival distributions of jam durations (Fig.~\ref{fig:survival}\textbf{e}) are compatible with power-law decays with exponents smaller than 2 in absolute value for $N=24$ and $N=30$. In this case, the power laws span over almost two orders of magnitude over the time axis. From this observation, we can represent the average duration of jams $\langle d \rangle$ vs. $N$, obtaining three well differentiated regions (Fig.~\ref{fig:survival}\textbf{f}). For diluted systems ($N<15$) jams are sporadic, last for very short periods of time, and are typically caused by just one robot moving a little bit more slowly than the one behind it; therefore $\langle d \rangle \approx 0$. As $N$ increases jams begin to appear in earnest, leading to finite values of $\langle d \rangle$. Importantly, jam persistence gives rise to the characteristic emergence of stop-and-go waves. Finally, for still larger values of $N$, jam duration diverges, in the sense that the calculation of its average depends on the time window employed (the longer the measurement, the higher the average duration). In Fig.~\ref{fig:survival}\textbf{f} this is represented by a dot with a dashed line in the vertical direction, indicating that the average value calculated would tend to infinity for infinitely long measurements. More importantly, this behavior can be associated with a totally congested scenario in which the jam duration and size percolate through the whole system.

From these experimental results, a concern may arise regarding whether the transition to totally congested flow is caused by finite-size effects. To address this issue, we conducted numerical simulations using a variation of the Nagel-Schreckenberg cellular automaton model for freeway traffic \cite{NagelSchreckenberg}. We implemented a one-dimensional array of $L = 333$ sites with periodic boundary conditions where we evenly distribute $N_S$ simulated agents. Agents are initially at rest and start moving, increasing their speed by one unit per simulation step until reaching a maximum speed $V_{Smax}$ (different runs were implemented with $V_{Smax}= 2$ and $4$ sites/step). To introduce some degree of randomness in speed, at each time step each agent has a probability $p=0.2$ to reduce it by one unit. Moreover, agents stop to avoid collisions with preceding ones; and ghosting overtakes are also forbidden. Each simulation run lasts for 200 steps and 10 repetitions are implemented for each set of parameters (see SM for additional details).

First, we tried to reproduce the behavior of macroscopic quantities reported in Fig.~\ref{fig:dia}\textbf{a-c}. Importantly, to reach this goal we needed to implement a feature accounting for the detection distance at which the robots stop moving. This was accomplished by defining a detection threshold of $d_d=9$ cells; when the preceding agent is less than $d_d$ cells away, the simulated agent stop. This introduces randomness in the stopping distance (as it ranges from $d_d=9$ to $d_d-V_{Smax}$), a phenomenon also observed in the experiments. As evidenced by the lines in Figs.~\ref{fig:dia}\textbf{a-c}, the agreement between simulations and experiments is quite good. Building on this, we extended the system size to $10^4$ sites and computed the distributions of jam duration $d_S$ for different number of simulated agents $N_S$ (Fig. \ref{fig:simuls}\textbf{a}). Remarkably, we qualitatively reproduce the trends observed experimentally, with the transition to a totally congested scenario (exponent below 1 in the survival function) taking place at about $N_S=750$. The representation of the average jam duration $\langle d_S \rangle$ vs. $N_S$ (Fig. \ref{fig:simuls}\textbf{b}) also exhibits a behavior similar to the experimental one, although the transitions occur at higher values of $N_S$. Interestingly, if we calculate how many robots in the experiment would correspond to the densities simulated with $N_S$ and $10^4$ sites ($N_E$, see top x axis in Fig. \ref{fig:simuls}\textbf{b}), we achieve a reasonable agreement with the experimental results. These findings confirm that the existence of three distinct phases (flowing, congested, and totally congested) is an intrinsic hallmark of the studied system that emerges independently of its size.

In conclusion, we report experimental and numerical results of the single file motion of a robot swarm where we reproduce the transition from a free flow phase to a congested one, characterized by intermittent flow and the emergence of stop-and-go waves due to the development of jams. Moreover, the statistical analysis of the jam size and duration enables to discern a transition to a totally congested scenario wherein jams percolate over the entire system, and their average duration diverges. Beyond the interesting features reported, this work showcases the great potential of robotic systems to reproduce complex collective behavior observed in various fields, including active matter and pedestrian dynamics. Among the main advantages of this system, we highlight the temporal stability of the experimental conditions, the high degree of control over various parameters (including speed, force, reaction time, and interparticle distance  \cite{interparticeldistancing}), and the versatility in terms of the situations that can be implemented thanks to the robot programmable capabilities. A near future challenge involves extending the experiments to bidimensional scenarios and enhancing control over robot-robot communication to account not only for short range, but also for long range interactions.

\begin{acknowledgments}
We specially acknowledge L. F. Urrea for technical help. This work has been funded by Grant No. PID2020-114839GB-I00 supported by MCIN/AEI/10.13039/501100011033.
\end{acknowledgments}

\end{document}

\title{Supplemental Material for `Single file motion of robot swarms'}

\author{Laciel Alonso-Llanes}

\author{Angel Garcimart\'in}
\email{angel@unav.es}

\author{Iker Zuriguel}
\affiliation{Departamento de F\'{i}sica, Facultad de Ciencias, Universidad de Navarra, E-31080 Pamplona, Spain}

\maketitle

In this supplemental material we provide detailed information about the robot characteristics, the software used, and their motion. We also describe the Nagel-Schreckenberg cellular automaton model we implemented, specifying the modifications with respect to the original one.

\section*{Experiments}

\subsection{Robot features}

\begin{figure}[ht]
\includegraphics[width=0.8\columnwidth]{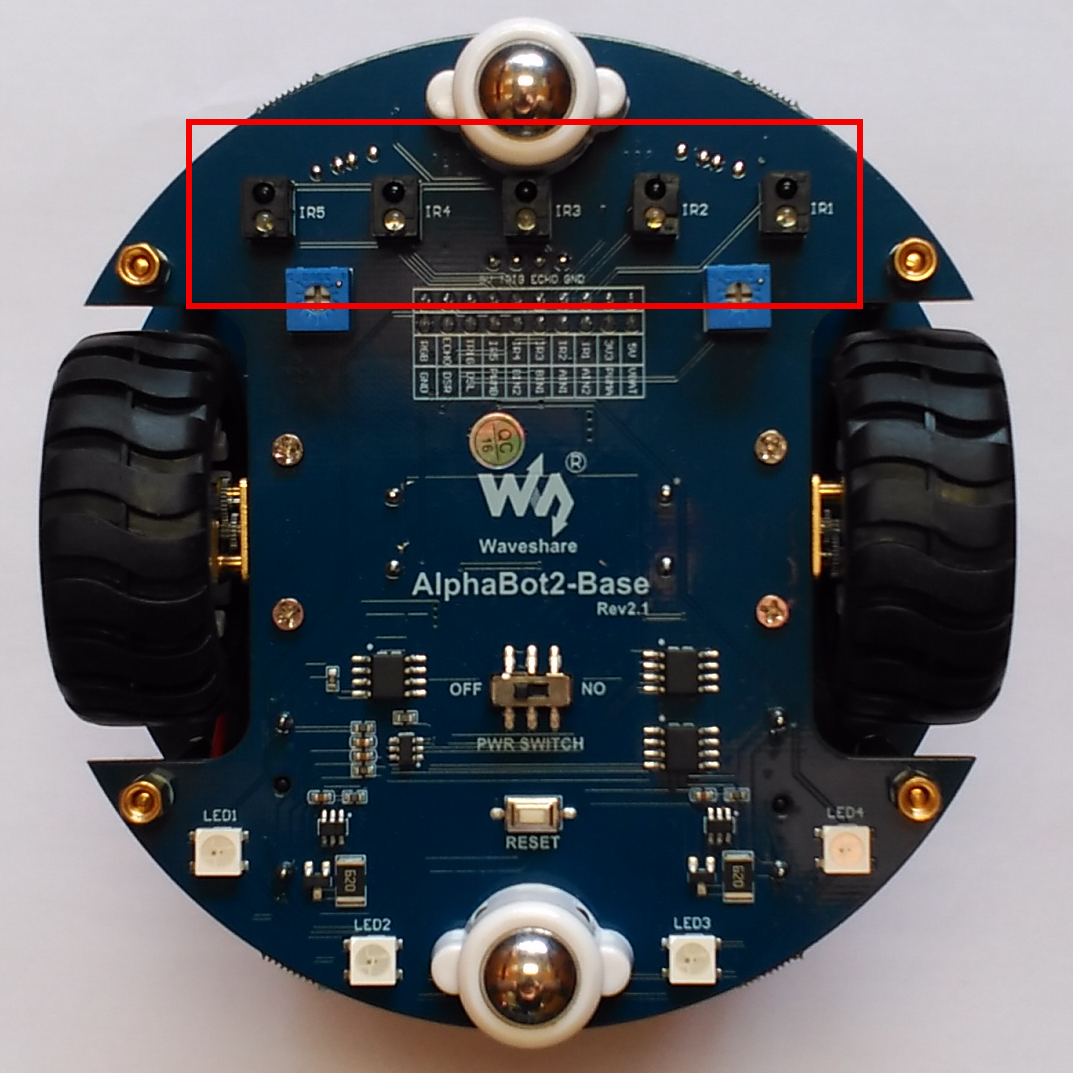}
\caption{\label{fig:rbott} Photograph of the bottom of the robot, highlighting the line array of infrared sensors with a red box. The two wheels on the right and left sides are individually powered by electric motors. Additionally, two metal spheres in white casings (top and bottom), designed for free rotation, contribute to maintain the robot balance.}
\end{figure}

The robots employed in the study were constructed using a commercially available platform known as Alphabot2-Pi, manufactured by Waveshare \cite{waveshare}. These robots are built on an 11 cm diameter chassis equipped with two driven wheels at the sides and two idle wheels at the front and rear to ensure stability (Fig.~\ref{fig:rbott}). Each wheel is powered by an independent DC motor. The chassis accommodates two batteries for power supply and a Raspberry Pi 3B+ nanocomputer with interfaces to various sensors and devices, including control over the wheel motors. The speed is regulated by adjusting the duty cycle (expressed as a percentage of the maximum power) of the electric motors. Since the Raspberry Pi is accessible via Wi-Fi, all programming tasks can be performed remotely.

In the current experiment, we used two sets of sensors integrated into the Alphabot2-Pi. The first set comprises a line array consisting of five infrared sensors at the bottom, arranged perpendicular to the central axis and spaced at 1.27 cm intervals (see red box in Fig.~\ref{fig:rbott}). With these sensors, the robots are able to follow a dark line on a white background. To this end, the robots use a software PID (Proportional-Integral-Derivative) control loop that adjusts the duty cycle of each motor after the imbalance among the sensor readings.

The second set of sensors comprises two tunable IR proximity sensors positioned at the mid-front, oriented horizontally to detect obstacles along the path. To enhance detectability, we constructed a circular shield for each robot, forming a 2.5 cm high band around it (with the front part open, where the IR sensors are located), covered with reflective tape. This design ensures that robots are easily detected by others, at a uniform distance, irrespective of the approach angle. The sensitivity of the sensors was then adjusted so that the detection distance was set to 15 cm. Both elements—the white shield and the proximity sensors—are visible in Fig. 1a in the article.

\subsection{Velocities and braking}

\begin{figure}[ht]
\includegraphics[width=\columnwidth]{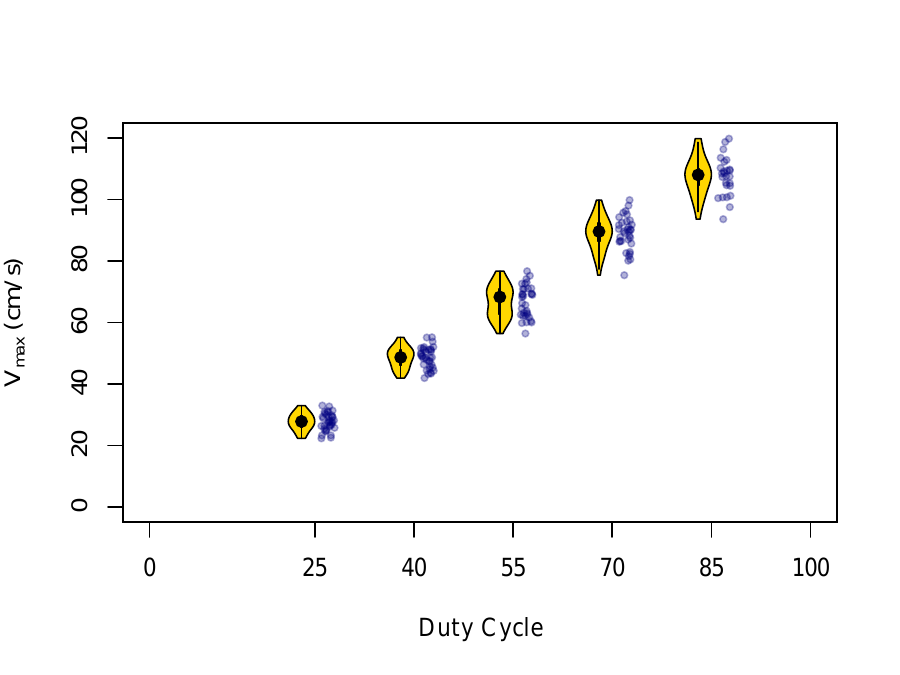}
\caption{\label{fig:SF1} The mean free speed of 26 individual robots is shown as a function of the nominal duty cycle, represented by blue dots. The distribution of the ensemble is visualized through a violin plot. A discrete set of duty cycles, corresponding to the values indicated in the labels, was employed for this representation.}
\end{figure}

It is important to note that even when the duty cycle of the motors is set to a fixed nominal value, the free speed of a single robot (i.e., the speed when a single robot follows a line without any obstacle) undergoes slight variations over time. This occurs because the PID controller calculates and supplies some extra power to one wheel in order to follow the dark line. These variations are short lived and display a Gaussian distribution, with a standard deviation of less than 5\% of the average speed. Furthermore, the average free speed differs from one robot to another at the same nominal duty cycle. This variability among robots stems from minor disparities in the motors and bottom sensors. A variability of approximately 8\% has been observed (Fig.~\ref{fig:SF1}). The mean free speed averaged for the ensemble corresponding to a given duty cycle is denoted as $V_{max}$ in the article.

Special procedures must be taken so that the robot comes rapid to a halt when an obstacle is detected (meaning, when another robot is in front) in order to avoid collisions. To this end, a specific piece of code was incorporated into the program. Upon detection of an obstacle, the motors promptly set the wheels in reverse motion at 25\% of the nominal duty cycle for 0.2 seconds. This implementation allows the robot to come to a halt in approximately 9 cm for $V_{max}=69$ cm/s and at 5.5 cm for $V_{max}=28$ cm/s. These numbers have been obtained from the measurements of the detection distance (15 cm in both cases), and the separation measurement between consecutive stopped robots: for robots moving with $V_{max}=69$ cm/s, the average separation was of 17 cm, while for $V_{max}=28$ cm/s we measured an average of 20.5 cm. A more aggressive braking procedure may lead to the robot losing balance or skidding, resulting in less consistent braking distances.

\subsection{Experimental procedure}

The experimental procedure is conducted as follows: the robots are arranged at uniform intervals on a circular black line, printed on paper, measuring 2.54 cm in width and 2 m in diameter. Using the Cluster Secure Shell (cssh) administration tool, all the robots are simultaneously accessed and a Python program is launched on them at the same time, so that they start moving in synchrony. The program sets the robots into motion with the same nominal duty cycle value for all. As they move forward, following the black line, they stop upon detecting another robot in front (the detection distance is 15 cm from the IR sensor to the obstacle). The robots remain stopped until a negative detection occurs, at which point they resume movement, accelerating to the maximum speed at the power indicated by the duty cycle. As said in the main manuscript and shown in Fig. 1c there, it is important to remark that the acceleration to $V_{max}$ is slower than the braking process.

If all the robots move at precisely the same speed, no robot would stop, as the separation among them would remain strictly constant over time. It is only due to small variations in speed that robots can approach or separate from the preceding one. Therefore, the initial conditions implemented in this work (robots arranged at uniform intervals) are the least likely to create jams (or, conversely, they are optimal for achieving a continuous flow).

\begin{figure}[ht]
    \includegraphics[width=\columnwidth]{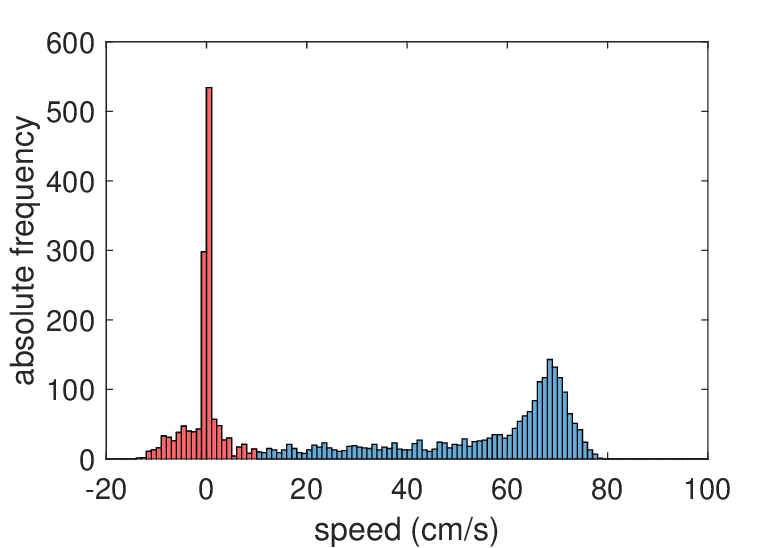}
    \caption{\label{fig:SFv1} Histogram of speed measurements for a single robot using a bin size of $1$ cm/s. A speed threshold of 10 cm/s is used to define two states: stopped (\textit{red}) and running (\textit{blue}). In this instance, the robot was part of a group of 20 moving along the circular line at a nominal speed of 69 cm/s. The histogram also illustrates the time that the robot spent in a stopped and moving state.}
\end{figure}

The typical duration of an experimental run is 3 minutes, during which the robots are recorded from above using a video camera with a resolution of $2600 \times 2600$ pixels at a rate of 20 frames per second. Subsequently, the recorded video is processed frame by frame to detect the centroid of the red square on top of each robot. The positions of individual robots are linked to form trajectories. Speed is calculated at each frame by subtracting the previous position and dividing by the sampling time (0.05 s). The histogram of speeds measured for a single robot is illustrated in Fig.~\ref{fig:SFv1}, displaying a clear bimodal distribution. One peak corresponds to the stopped robot, and another peak represents moments when the robot is in motion. A threshold of 10 cm/s is set to categorize all measured speeds into two situations: a stopped robot (\textit{red}) and a running robot (\textit{blue}). Other strategies, such as setting a variable threshold as a percentage of $V_{max}$, were tested without notable changes in the results. Note that the peak corresponding to the robot moving (\textit{blue}) is wider than it would be if the robot were alone, as it includes braking and acceleration events. The occurrence of some \textit{negative} speeds in the region corresponding to the robot stopped (\textit{red}) is due to the top of the robot swinging backward after braking.

\section*{Numerical Simulations}

\begin{figure*}[t]
    \centering
    \includegraphics[width=0.7\textwidth]{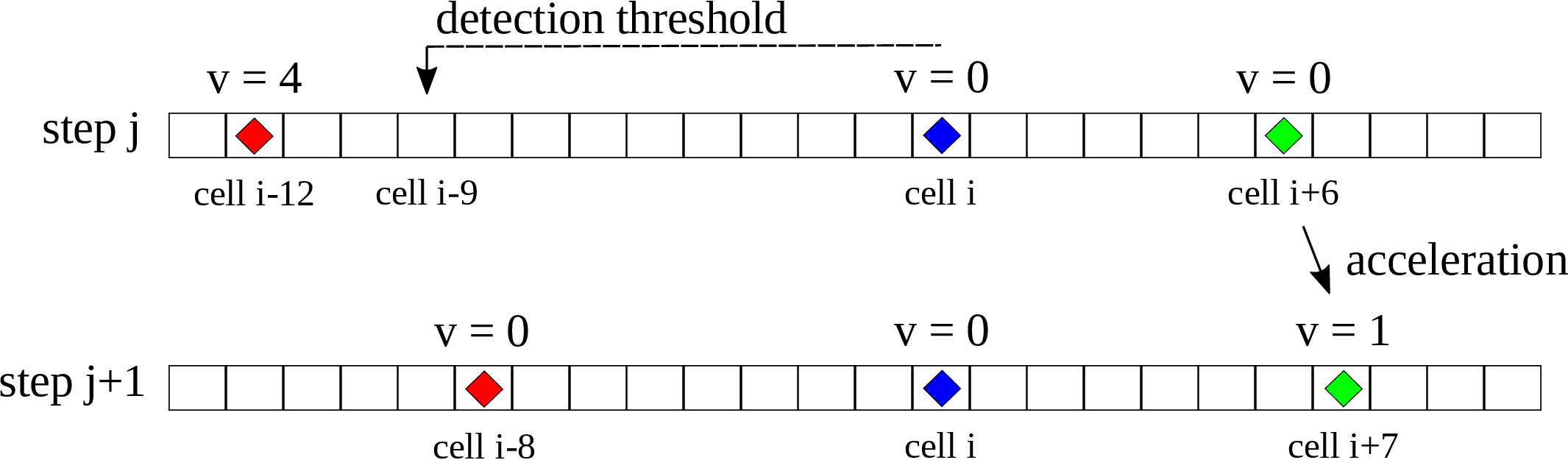}
    \caption{\label{fig:SD} Illustration of two consecutive simulation steps depicting some cells for a scenario in which a red agent comes to a halt upon detecting the blue agent in front of it. In the step $j$ the red agent is moving freely at its maximum speed ($v=4$). Then, in step $j+1$ the red agent moves to cell $i-8$, detects the blue one at a distance smaller than $d_d=9$ and stops. The diagram also includes the acceleration of the green agent which has $v=0$ in step $j$ and increases it to $v=1$ in step $j+1$ as there are no agents in front of it. Note that the blue agent does not accelerate since it is detecting the green one at a distance smaller than 9 sites.}
\end{figure*}

The simulations are implemented with a cellular automaton model that operates on a one-dimensional array of $L$ sites with periodic boundary conditions. Each site in this array can be either occupied by an agent or left empty. These agents have integer speeds ranging from $0$ to $V_{S\text{max}}$, representing their maximum achievable speed during acceleration. The initial positions of the agents are arranged at regular intervals, taking into account the periodic boundary conditions to ensure a continuous environment for their interactions.

A crucial factor influencing the decision-making process of each agent is the detection distance, \( d_d \), representing the range within which an agent can detect the preceding robot. Additionally, the slowdown probability (\( p \)) parameter introduces variability by influencing random slowdown events for each agent.

For an arbitrary configuration, one update of the system consists of the following consecutive steps, which are performed in parallel for all agents (see Fig \ref{fig:SD} for an illustration):

\begin{itemize}

    \item \textbf{Acceleration:} If the current speed of the agent is less than the maximum \( V_{Smax} \), its speed is increased by one unit.

    \item \textbf{Braking:} The speed of an agent is set to zero if the distance from the one in front is less than or equal to the detection distance: $d_d=9$ cells in this case.

    \item \textbf{Random Slowdown:} At each step, a random number between 0 and 1 is generated for every agent; if the number is smaller than $p$, the agent decreases the speed by 1 unit, provided it is not already zero.

    \item \textbf{Motion:} The robot positions are updated by adding to each robot current position the number of cells corresponding to its speed.

\end{itemize}

This routine governs the dynamic behavior of the agents, ensuring the avoidance of collisions and `ghosting' overtakes, while enabling controlled acceleration and braking in response to neighboring agents and random perturbations. It ensures an acceleration to $V_{max}$ slower than the braking process. For each conditions explored the program is run for 10 independent realizations for a prescribed number of steps $T_S$. The code used is available in \cite{GITHUBLINK}.

In our first approach, we tried to replicate the experimental findings reported in Fig. 2(a) of the main text. To this end $T_S=200$ and the road length was 333 cells. The scenario of $V_{max}=69$ cm/s is simulated with a maximum speed of $V_{Smax}=4$ cells/step, while the scenario of $V_{max}=28$ cm/s uses $V_{Smax}=2$ cells/step. Importantly, trying to account for the experimental observation that fast robots stop at shorter distances than slow ones, we adopt a detection distance $d_d=9$ cells for $V_{Smax}=4$ cells/step, and $d_d=11$ cells for $V_{Smax}=2$ cells/step. Later on, we tried to mimic larger scenarios by means of a road length of $10^4$ cells and $5600$ time steps (Fig. 4 of the main text). In this case, we focused on the scenario of $V_{Smax}=4$ cells/step, analyzing the role of $N$ and keeping the rest of parameters identical.

\subsection*{Time and space equivalence}

To establish temporal and distance equivalences among experiments and simulations, we used a calibration process based on experimental data. The size of a cell was determined by equating the real distance at which robots stop from each other (from center to center) to the detection distance implemented in the simulations. For robots moving with $V_{max}=69$ cm/s, we measured that the separation between consecutive stopped robots was of 17 cm, while for $V_{max}=28$ cm/s we obtained 20.5 cm. Consequently, a cell was defined as equivalent to 1.89 cm ($17/9$) and 1.86 cm ($20.5/11$) for $V_{max}=69$ cm/s and $V_{max}=28$, respectively.

Once this spatial relationship is determined, the temporal conversion is straightforwardly calculated from the ratio among experimental and numerical robots average speeds. This ratio is multiplied by the cell size in centimeters obtaining the relationship between simulation time steps and real-world seconds. The computed figures are 0.11 and 0.14 s/step for $V_{max}=69$ and $28$ cm/s, respectively. Note that the difference among these figures comes from the fact that the ratio among the two simulated velocities is 2, whereas the ratio among the two experimental ones is slightly higher (2.4).

Obtaining temporal and distance equivalences allowed us to compare numerical results with experimental ones (Fig. 2(a-c) of the main manuscript), as well as maintain consistency in the temporal aspects of the simulations across varying scales. Indeed, for the extended simulations reported in Fig. 4, the equivalent duration and road length obtained are 616 s and 189 m respectively.

\section*{Jam definition}

To identify and quantify traffic jams in the simulations, a definition must be provided first to decide whether or not two nearby stopped robots belong in the same jam. In order to do this, a binary matrix is employed defining a neighborhood of cells, both in space and time. If a stopped robot placed at the center of this matrix has another stopped robot in the neighborhood at the positions and times indicated with 1 in the connectivity matrix, then both belong in the same jam. Subsequently, a simple union-find algorithm using this matrix as a kernel is applied to establish connections among clusters of ones in the space-time cell mesh of the whole simulation. The following Connectivity Matrix has been used, where the horizontal index represent space and the vertical represents time:

\[\text{Jam Connectivity Matrix} =
\begin{bmatrix}
    0 & 0 & 1 \\
    0 & 1 & 1 \\
    1 & \text{x} & 0 \\
    0 & 0 & 0 \\
    0 & 0 & 0 \\
\end{bmatrix}
\]

Importantly, we must take into account the periodic boundary conditions implemented in our system. This is performed considering that the agent $N$ is followed by agent 1. Consequently, a jam or cluster may extend across the matrix border, establishing connectivity between the last and first columns in the sequence. Having said that, we remark that the maximum reachable cluster size is $N$, no matter if an agent participates in a given jam more than once (i.e. if the agent abandons the jam and joins it back).

For the case of experiments, a similar procedure is followed. After our observations, the connectivity matrix extends backwards in time for 0.5 s to take into account the reaction time of robots.